\documentclass{aa}  

\usepackage{graphicx}
\usepackage{txfonts}
\usepackage{lipsum}
\usepackage{subcaption}         
\usepackage{lscape}             
\usepackage{placeins}           
\usepackage{booktabs}
\usepackage{siunitx}
\usepackage{isotope}
                    
\usepackage[colorlinks=true,
            citecolor=blue,
            linkcolor=blue,
            urlcolor=blue]{hyperref}

\newcommand{\acr}[1]{\texttt{#1}}

\begin{document}

   \title{The origin of Potassium in Galactic chemical evolution}

   \subtitle{The effect of C--O shell mergers in massive stars}

%
%
%

   \author{M. Molero\inst{1,2,3}\corrauth{marta.molero@csfk.org}
        \and L. Boccioli\inst{1,4}\email{luca.boccioli@tu-darmstadt.de}
        \and A. Arcones\inst{1,5,6}\email{almudena.arcones@tu-darmstadt.de}
        }

   \institute{Institut f\"ur Kernphysik, Technische Universit\"at Darmstadt, Schlossgartenstr. 2, Darmstadt 64289, Germany
   \and Konkoly Observatory, HUN-REN Research Centre for Astronomy and Earth Sciences, H-1121 Budapest, Konkoly Thege M. út 15-17, Hungary
   \and CSFK, MTA Centre of Excellence, Budapest, Konkoly Thege Miklós út 15–17, H-1121, Hungary
   \and Department of Physics, University of California, Berkeley, CA 94720, USA
   \and 
        GSI Helmholtzzentrum f\"ur Schwerionenforschung GmbH, 64291 Darmstadt, Germany
    \and
        Max-Planck-Institut f\"ur Kernphysik, Saupfercheckweg 1, 69117 Heidelberg, Germany
}
   \date{Received September 30, 20XX}

 
  \abstract
   {The Galactic chemical evolution (GCE) of potassium (K) and of most of the other odd-Z elements remains poorly understood. Models generally underproduce the observed [K/Fe] over a wide range of metallicities, depending on the adopted massive star yields set. A convective-reactive phenomenon that has been proposed as an efficient environment for the nucleosynthesis of K is the merger between the oxygen (O) and carbon (C) shells that can occur in massive stars a few hours before core-collapse.}
   {A few different GCE models have been used to show that, for specific massive star yields, one can reconcile the tension between the underproduction of K seen in traditional GCE models and observations. Here, we re-analyze some of those massive star sets to investigate how large the contribution of C--O shell mergers is to the production of K, and what we can infer on the occurrence of these events from the Galactic evolution of K.} 
   {We analyse 194 pre-supernova models from three independent yield sets, which together cover a wide range of initial masses, metallicities and rotational velocities, and include both single and binary-stripped stars. We identify the progenitors that undergo a C--O shell merger and compute the GCE models with and without their contribution. We then decompose the K production by progenitor mass, both for single stellar generations and over the whole Galactic history.}
   {In all three sets, the enhanced K production can be traced back to a small number of C--O merger progenitors which eject up to $\sim$10--50 times more K than progenitors without C--O shell mergers of similar mass. Without the contributions from these selected progenitors, none of the models reproduce the observations, and [K/Fe] is sub-solar over most of the metallicity range. With the C--O merger progenitors, [K/Fe] is enhanced only within the metallicity window where mergers occur in the grid, which varies from one set to the other. Rotating progenitors happen to have more C--O shell mergers at [Fe/H]$\lesssim -1$, while non-rotating ones have more C--O shell mergers closer to solar metallicity. Since the binary-stripped progenitors and their single-star counterpart are only available at solar metallicity, the extension to lower metallicities has to be done manually, and therefore the occurrence of C--O shell mergers is only mildly dependent on [Fe/H]. We find that binarity itself plays a minor role in the production of K, which is produced in a similar amount also by a population of single stars only.}
   {The Galactic evolution of K traces the location of the C--O merger progenitors in the mass-metallicity plane. Since the K synthesised during a C--O merger is essentially primary, the observed nearly linear increase of [K/H] with [Fe/H] can be reproduced only if C--O shell mergers occur with a similar frequency at all metallicities. This suggests that K could potentially be used to infer the frequency of occurrence of C--O shell mergers in massive stars as a function of time.}


   \maketitle
   \nolinenumbers


\section{Introduction}

Galactic chemical evolution (GCE) models are powerful tools to investigate the origin and the dispersion of chemical species in the interstellar medium (ISM) of galaxies. Among their ingredients, stellar yields are often the dominant source of uncertainty, as modifications in the yields prescription change the predicted abundance patterns more than any reasonable variation of the star formation history or of the galaxy formation and evolution prescriptions (\citealp{Romano2010, Liang2023}). The link between stellar nucleosynthesis calculations and GCE models can therefore be exploited to test the assumptions made in stellar evolution and nucleosynthesis against the abundances measured in large populations of stars. The systematic deviations that emerge from this comparison can then be used to identify the physical ingredients to which the observations are most sensitive: stellar rotation and rotation-induced mixing (\citealp{Limongi2018_PresEvolExplNuclRotaMassStar, Prantzos2018}), the explosion energy and location of the mass cut (\citealp{Nomoto2013, Jost2025}), the $\nu$-process (\citealp{Kobayashi2011Fluorine}), or the convective-reactive episodes that can take place in the last stages before the explosion (\citealp{Ritter2018_ConvReacNuclScClPproIsot, Cote2020}).

Odd-$\rm Z$ elements such as phosphorus (P, $\mathrm{Z = 15}$), chlorine (Cl, $\mathrm{Z = 17}$), potassium (K, $\mathrm{Z = 19}$) and scandium (Sc, $\mathrm{Z = 21}$) are potentially well suited to this purpose, since their astrophysical origin is still a matter of debate. K, in particular, whose dominant isotope $\rm ^{39}K$ is predicted to be synthesized by hydrostatic and explosive oxygen burning in the inner regions of massive stars (\citealp{Woosley1995_EvolExplMassStar2ExplHydr}), is often underproduced by GCE models. The metallicity range over which the discrepancy appears varies depending on the adopted set of massive star yields, hence the predicted [K/Fe] is found to be underproduced at all metallicities (\citealp{Kobayashi2020_OrigElemCarbUran}), only at low-to-intermediate [Fe/H] (\citealp{Ritter2018_ConvReacNuclScClPproIsot}), or only at high [Fe/H] (\citealp{Prantzos2018}). 

\citet{Ritter2018_ConvReacNuclScClPproIsot} attributed the ability of their models to enhance the production of K towards solar metallicity to the synthesis of this element (and of the other odd-Z species) in convective-reactive episodes during the pre-supernova evolution, and specifically to the C--O shell mergers that develop in the $12$, $15$ and $20\ \rm M_\odot$ models at the highest metallicities of the progenitor grid, $\rm Z = 0.01$ and $\rm 0.02$. During a C--O shell merger, when the entropy barrier between the O and C shells disappears, material in the C shell is ingested and mixed all the way to the bottom of the O-shell \citep{Ritter2018_ConvReacNuclScClPproIsot,Roberti2025}. Therefore, fresh \isotope[16]{O} as well as some \isotope[12]{C} left over from C-burning, are added into the O-burning region, triggering an O-burning event, mostly powered by $\isotope[16]{O}+\isotope[16]{O}$ and $\isotope[12]{C}+\isotope[12]{C}$ reactions. This releases protons, $\alpha$ particles as well as some neutrons, which in turn are responsible for producing odd-Z elements, such as K, as well as many others \citep{Roberti2025_SPArBurnProtCaptPoweCarbShel}. In principle, the innermost part of this region could be reprocessed by the shock once the explosion sets in, thereby erasing this nucleosynthetic signature. However, it has been shown that the abundances in this region are largely preserved even after the shock has swept the C--O shell \citep{Boccioli2026_ProdHeavAEle44TiCasCompAbun}. All of these C--O merger events occur in 1D stellar evolution codes, but there exist a few works where the shell merger event was simulated in three dimensions for a few hours, with different methods and choices for nuclear burning \citep{Andrassy2020_3DHydrSimuIngeConvShel,Rizzuti2023_3DStelEvolHydrSimuCompBurn,Rizzuti2024_ShelMergLateStagMassStarEvol}. However, the nucleosynthesis can only be reliably calculated from 1D models. Indeed, some of the aforementioned 3D simulations either have a small network coupled to the hydro, insufficient to fully characterize the nucleosynthesis, or don't have one at all, and instead use 3D-informed mixing coefficients to post-process the underlying 1D stellar model. Given the large impact of C--O shell mergers on the production of select elements, one can use GCE models to place constraints on the frequency of these convective-reactive events based on their ability to reproduce the observed Galactic evolution of the element in question.

In this manuscript, we show that the few GCE models that show an enhanced production of K (even if only over a limited metallicity range) can in all cases be traced back to a small number of specific progenitor models that underwent a C--O shell merger. To this purpose, we analyse a total of $\rm 194$ progenitors spanning different initial progenitor masses, metallicities and rotation velocities. Three main independent yield sets are considered: the rotating massive star models of \citet{Limongi2018_PresEvolExplNuclRotaMassStar} (hereafter \acr{LC18}), the single and binary-stripped models of \citet{Farmer2023_NuclBinaStar} (hereafter \acr{F23}), and the non-rotating, single-star NuGrid models of \citet{Ritter2018_NuGrStelDataSetIIStelYiel} (hereafter \acr{R18}). We run GCE models for each of these sets in order to show the sensitivity of K production to the occurrence of C--O shell mergers in a given stellar population.

The paper is structured as follows. In Sect. \ref{sec: C--O shell mergers} we summarise the physics of the C--O shell mergers, present the criterion adopted to identify them in the pre-supernova models and list the C--O merger of each set. In Sect. \ref{sec: nucleosynthesis prescriptions} we describe the chemical evolution model and the three sets of massive star yields and how they are included in the model, together with the other stellar sources. In Sect. \ref{sec: results}, we compare the predicted evolution of K with the observations, with and without the contribution of the C--O merger, and we identify the progenitor masses that dominate the Galactic K budget in each set, while in Sect. \ref{sec: caveats} we discuss the main caveats of the approach. Our conclusions are summarised in Sect. \ref{sec: conclusions}.


\section{Carbon-oxygen shell mergers}
\label{sec: C--O shell mergers}

In the last phases of the life of a massive star, the region above the contracting Fe core is stratified in nuclear burning shells. The innermost region is the Si-burning shell, overlaid by the convective O-burning shell, and then by the convective C-burning shell and by its C- and Ne-rich ashes. The O and C convective shells are usually separated by a stable layer, but this configuration is not always preserved. Under specific conditions, the O shell can grow outwards by entraining the material at the outer edge of the convective zone, until the C- and O- shells merge into a single convective region. This is what is referred to as a C--O shell merger, and it happens a few hours before core-collapse. The exact cause and dynamics of this process are still unclear, and more 1D and 3D studies are needed in order to fully understand this phenomenon. Generally, 3D simulations show much larger entrainment rates during shell-burning events   (\citealp{Meakin2007_TurbConvStelInte1HydrSimu, Jones2017_IdeaHydrSimuTurbOxygShelConv, Andrassy2020_3DHydrSimuIngeConvShel, Rizzuti2023_3DStelEvolHydrSimuCompBurn, Rizzuti2024_ShelMergLateStagMassStarEvol}) compared to simpler overshooting and convective-boundary mixing (CBM) models in 1D. This can therefore cause the O shell to quickly grow until it eventually reaches the C shell, triggering a shell merger event. At the same time, even 1D models computed without any CBM can exhibit fully developed C--O shell mergers \citep{Roberti2025_OccuImpaCarbShelMergMassStar}. It should be noted, finally, that this process occurs just a few hours before collapse, when nuclear burning is an exceptionally stiff, non-linear problem in the interior, where even very large and detailed nuclear networks in 1D become unstable \citep{Renzo2024_ProgStarCalcSmalReacNetwShou, Grichener2025_NuclNeurNetwEmulLateBurnStag}, whereas 3D simulations often adopt very small nuclear networks, and sometimes none at all.

Once a C--O shell merger event is triggered, fresh C- and especially O-rich fuel are carried down to the base of the O shell, where the temperature is of the order $2 - 3\ \rm GK$, typically enough to activate the heavy-ion fusion reactions $\rm ^{12}C + ^{12}C$, $\rm ^{12}C + ^{16}O$ and $\rm ^{16}O + ^{16}O$, which release a large number of light particles ($p$, $\alpha$, $n$). These are captured by the products of O burning, and the odd-Z isotopes $\rm ^{31}P$, $\rm ^{35}Cl$, $\rm ^{39}K$ and $\rm ^{45}Sc$ are synthesised mainly through a chain of proton and $\alpha$ captures on $\rm ^{34}S$, such as $\rm ^{34}S(\alpha, \gamma) ^{38}Ar(p, \gamma)^{39}K$. This production also appear to be primary \citep{Roberti2025}, as the odd-Z elements are built from the C, Ne and O of the star itself, and not from the metals it was born with, so that the process should operate pretty much unchanged at any metallicity. Therefore, since the nucleosynthesis itself should be largely independent of metallicity, what drives the change in odd-Z element production throughout the history of the Galaxy is how frequent the shell-merger events are as a function of metallicity, which remains an open question.

\begin{figure*}
    \centering
    \includegraphics[width=\textwidth]{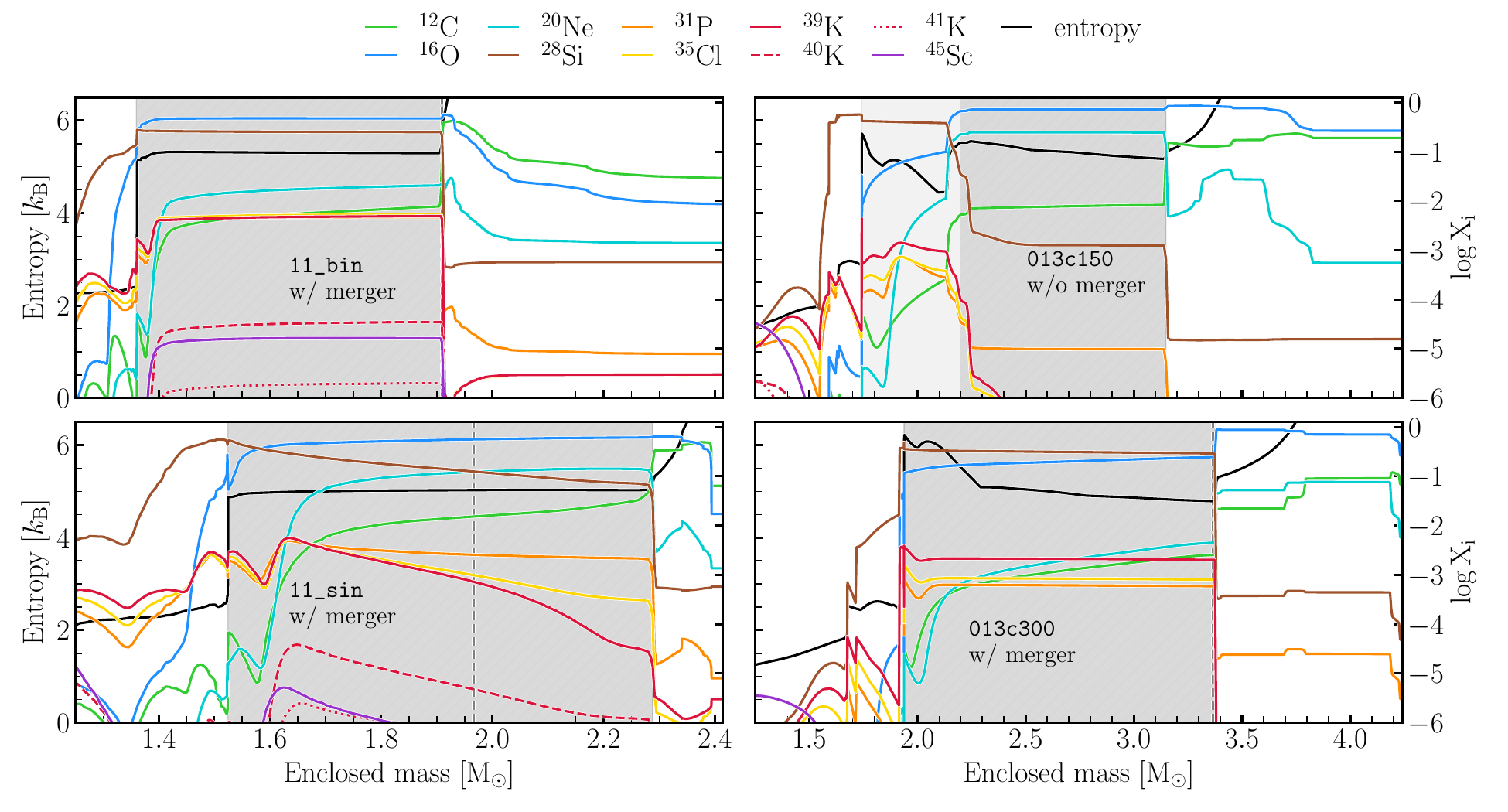}
    \caption{Examples of compositional and entropy profiles of pre-supernova progenitors at the onset of core collapse with and without a C--O shell merger. On the left panels we show the $11 M_\odot$ single (bottom) and binary (top) progenitors from \acr{F23}, both with a C--O merger. On the right panels we show the $13 M_\odot$ progenitors at [Fe/H]$=-2$ with initial rotational velocities of 150 (top) and 300 $\rm km\ s^{-1}$, the first without and the second with a C--O merger. Entropies are shown in black, whereas the mass fractions $\rm X_i$ are shown in different colours and line styles on a logarithmic scale. The light grey shaded region represents the O-burning shell, whereas the dark grey shaded region represents the C shell. The top right panel is the only one where these two regions are clearly distinct, whereas in all other cases they coincide, since they all experienced C--O shell mergers. The vertical dashed line in the bottom left plot marks the location of the C--O shell where the mass fraction of \isotope[28]{Si} falls below \isotope[20]{Ne}, defining the fraction of C--O merger completeness shown in Tab.~\ref{tab:prog_details}.}
    \label{fig:StarStructures}
\end{figure*}

Since most sets of massive stars publicly available do not provide detailed snapshots of the pre-collapse evolution, but rather only the last one before core-collapse, identifying a C--O merger is non-trivial. Here, we flag a C--O shell merger when, in the combined O and C shells (dark grey band in Fig.~\ref{fig:StarStructures}), we find that the mass-averaged fraction of \isotope[28]{Si} is larger than that of \isotope[20]{Ne}. We list the models that satisfy this criterion in Tab.~\ref{tab:prog_details}, along with the merged fraction of the two shells, defined as the fraction of the shell in which \isotope[28]{Si} exceeds \isotope[20]{Ne}, and other relevant quantities. However, a more systematic analysis will be presented in a forthcoming paper. 
Fig.~\ref{fig:StarStructures} shows pre-supernova profiles for two $11\ \rm M_\odot$ models (a binary-stripped and a single star) from \acr{F23} and two $13\ \rm M_\odot$ models from \acr{LC18}. In the $\rm 11\ M_\odot$ binary-stripped model, the O and C shells are fully mixed, as entropy and composition are flat across the whole region, and \isotope[39]{K} is enhanced to $\sim5\times10^{-3}$ together with P and Cl, whereas Sc is about $\sim 10^{-5}$. The $\rm 11\ M_\odot$ single-star model also exhibits a C--O merger, but there, despite the flat entropy profile, the composition of some isotopes (like \isotope[39]{K}) has a negative gradient, a sign that perhaps the merger has not yet fully mixed the zone. This is what we define to be an incomplete merger (i.e. with a  $58 \%$ ``completeness" in Tab.~\ref{tab:prog_details}). The other complete C--O merger shown in the figure is the $\rm 13\ M_\odot$ model at $\rm [Fe/H] = -2$ with a $300~\rm km\ s^{-1}$ initial rotational velocity from \acr{LC18}. Being a lower-metallicity, fast-rotating star, it achieves larger CO and Fe cores. The C--O merger therefore has a much larger extension in mass, and O is burned much more efficiently, reaching an average mass fraction in the shell of $\sim 0.2$, versus $\sim 0.4-0.45$ in the two $11 M_\odot$ models. The more significant difference regarding the nucleosynthesis is in the production of \isotope[45]{Sc}. In the \texttt{013c300} model, the mass fraction of \isotope[45]{Sc} is $\sim 10^{-6}$, just below the scale shown, i.e., a factor of 10 smaller than for the two $11\ \rm  M_\odot$ models. This is a general feature for all C--O merger progenitors found in \acr{LC18} compared to the ones found in both \acr{F23} and \acr{R18}, where the mass fraction \isotope[45]{Sc} is usually around $10^{-5}$. Since we do not focus on Sc in this work, we simply comment that this might be due to different reaction rates and mixing prescriptions adopted, or perhaps to other rotational mixing effets in \acr{LC18}, which would therefore avoid potential pile-up effects. Finally, the $\rm 13\ M_\odot$ model at $\rm [Fe/H] = -2$ with a $300~\rm km\ s^{-1}$ initial rotational velocity from \acr{LC18} did not experience a C--O merger, and one can clearly see the difference in the lack of Si, K, P, Cl, and Sc in the dark grey shaded region, alongside the sharp entropy jump separating the two shells.

As an example, in App. \ref{sec: Kippenhahn diagrams} we describe the evolution of the $\rm 11\ M_\odot$ binary-stripped and of the $\rm 11\ M_\odot$ single models from \acr{F23}, reconstructed from the published \texttt{MESA} histories.

\begin{table}[!t]
\caption{List of the progenitor models that undergo a C--O shell merger and their properties.}
\label{tab:prog_details}
\centering
\small
\setlength{\tabcolsep}{4pt}
\sisetup{table-number-alignment = center}
\begin{tabular}{
  l
  S[table-format=1.2]
  S[table-format=1.2]
  S[table-format=1.2]
  S[table-format=1.2]
  S[table-format=1.2]
  S[table-format=2.2]
}
\toprule
Model & {Complete} & {$M^{\rm shell}_{\rm C-O}$} & {$M_{\rm CO}$} & {$M_{\rm He}$} & {X$(\isotope[12]{C})$} & {$M(\isotope[39]{K})$} \\{\vspace{1pt}}
 &   {[\%]} & {[$\rm M_\odot$]} & {[$\rm M_\odot$]} & {[$\rm M_\odot$]} &  & {[$10^{-3}$\,$\rm M_\odot$]} \\
\midrule
\texttt{013a300} & 57.20 & 1.36 & 3.67 & 5.30 & 0.15 & 1.61 \\
\texttt{020a150} & 52.85 & 2.02 & 5.51 & 7.63 & 0.21 & 5.18 \\
\texttt{013b150} & 63.12 & 1.29 & 3.78 & 5.10 & 0.13 & 2.27 \\
\texttt{013b300} & 100.00 & 1.44 & 4.31 & 5.83 & 0.03 & 3.41 \\
\texttt{015b300} & 100.00 & 1.99 & 4.86 & 6.55 & 0.06 & 4.53 \\
\texttt{013c300} & 100.00 & 1.44 & 4.20 & 5.70 & 0.02 & 3.91 \\
\texttt{015c150} & 85.51 & 1.58 & 4.49 & 5.80 & 0.13 & 2.44 \\
\texttt{015c300} & 100.00 & 1.68 & 4.69 & 6.51 & 0.04 & 4.24 \\
\texttt{020c300} & 35.36 & 2.15 & 7.09 & 9.93 & 0.00 & 3.16 \\
\texttt{015d000} & 71.19 & 0.96 & 3.27 & 5.20 & 0.21 & 2.61 \\
\texttt{015d150} & 100.00 & 1.15 & 4.11 & 5.99 & 0.10 & 1.72 \\
\texttt{015d300} & 30.99 & 1.39 & 4.53 & 6.24 & 0.14 & 14.72 \\
\addlinespace
\texttt{011\_sin} & 57.78 & 0.76 & 2.39 & 3.78 & 0.33 & 5.62 \\
\texttt{013\_sin} & 27.43 & 1.22 & 3.24 & 4.78 & 0.31 & 4.84 \\
\texttt{018\_sin} & 15.95 & 3.62 & 5.49 & 7.34 & 0.28 & 5.15 \\
\texttt{011\_bin} & 100.00 & 0.55 & 1.93 & 3.17 & 0.40 & 4.91 \\
\texttt{013\_bin} & 70.73 & 0.95 & 2.64 & 3.98 & 0.36 & 10.02 \\
\addlinespace
\texttt{m12z1e-2} & 41.95 & 0.60 & 2.22 & 3.63 & 0.21 & 0.82 \\
\texttt{m15z1e-2} & 10.87 & 1.52 & 3.24 & 4.89 & 0.25 & 0.58 \\
\texttt{m15z2e-2} & 100.00 & 1.44 & 3.07 & 4.71 & 0.24 & 1.50 \\
\texttt{m20z1e-2} & 2.48 & 3.62 & 5.18 & 7.13 & 0.22 & 0.70 \\
\bottomrule
\end{tabular}
\tablefoot{The first group shows models from \acr{LC18}. The naming convention indicates its ZAMS mass in $\rm M_\odot$ (first three characters), metallicity (a, b, c, and d stands for $\rm [Fe/H] = 0, -1, -2$ and $-3$, respectively), and initial rotational velocities in $\rm km\ s^{-1}$ (last three characters). The second group shows models from \acr{F23}, where the naming convention indicates its ZAMS mass in $\rm M_\odot$ (first three characters), and \texttt{sin} and \texttt{bin} denote the single and binary-stripped models. The third group shows models from \acr{R18}, where the naming convention indicates its ZAMS mass preceded by ``\texttt{m}" and its metallicity preceded by ``\texttt{z}". Other columns show, in order, the percentage of C--O merger completion, the mass of the merged C--O shell, the mass of the CO and He cores, the fraction of \isotope[12]{C} at He depletion, and the total mass of \isotope[39]{K} in the merged C--O shell.}
\end{table}

\section{Chemical evolution model}
\label{sec: model}
Since our aim is to isolate the effect of the C-O shell mergers on the Galactic evolution of K, rather than to reproduce in detail the K abundances of the different Galactic components, we adopt a standard GCE model of the Galactic disc and compare results that differ only in the massive star yields, with and without the contribution of the C-O merger progenitors. The predicted [K/Fe] and [K/H] are then compared with the general trends traced by the observations over the whole metallicity range, which also includes halo stars at low [Fe/H]. In particular, we use a one-infall multi-zone GCE model, based on the works of \citet{MatteucciFrancois1989, Chiappini2001, Grisoni2018}.

The disc is divided into concentric rings $1\ \rm kpc$ wide which evolve without exchanging matter. The results shown in this manuscript refer to the solar ring, at $R = 8\ \rm kpc$. Each ring is built up by the accretion of gas of primordial composition at an exponentially decreasing rate, $\propto e^{-t/\tau(R)}$, with an inside-out timescale $\tau(R) = 5 + 1.033\,(R\ {\rm kpc} - 8)$, normalised to reproduce the present day total surface mass density profile of the disc of \citet{NestiSalucci2013}.

The star formation rate follows a Schmidt-Kennicutt law, $\psi \propto \nu\,\sigma_{\rm gas}^{1.5}$, with an efficiency $\nu = 1\ \rm Gyr^{-1}$ in the solar ring. The instantaneous recycling approximation is relaxed, and each star restores its ejecta after its lifetime, taken from \citet{Kodama1997}. The initial mass function (IMF), $\varphi(m)$, is a \citet{Salpeter1955_LumiFuncStelEvol} power law, $\varphi(m) \propto m^{-\alpha}$ with $\alpha=2.35$, normalised to unit mass between $0.1\ \rm M_\odot$ and the most massive star of the adopted massive star yield set.

\subsection{The IMF of binary systems}
\label{sec: the imf of binary systems}

The \acr{F23} set (Sect. \ref{sec: nucleosynthesis prescriptions}) provides yields both for single stars and for stars that are stripped of their H-rich envelope by a companion through Roche-lobe overflow. These models span $11\leq M_1/\rm M_\odot \leq 45$ at solar metallicity and assumed a fixed mass ratio $q=M_2/M_1=0.8$. No yields are provided for the companions, i.e. the accretors. Therefore, when the stripped-star yields are used, the massive stars formed at each time step must be split into the primaries, which explode as stripped donors, and the secondaries, which evolve as accretors. We assume that all massive stars are members of binary systems ($f_{\rm bin} = 1$) and that every system has the same mass ratio $q=0.8$. Each star of mass $m$ is either the primary of a system with $M_1 = m$ or the secondary of a system whose primary has mass $M_1 = m/q$, so that
\begin{equation}
\label{eq:phi_sum}
    \varphi(m) = \varphi_{\rm p}(m) + \varphi_{\rm s}(m),
\end{equation}
where $\varphi_{\rm p}$ and $\varphi_{\rm s}$ are the mass functions of the primaries and of the secondaries, respectively. Since each primary has exactly one companion, the number of secondaries in the interval $[m, m + {\rm d}m]$ equals the number of primaries in $[m/q, (m + {\rm d}m)/q]$:
\begin{equation}
\label{eq:phi_s_def}
    \varphi_{\rm s}(m)\,{\rm d}m = \varphi_{\rm p}\!\left(\frac{m}{q}\right)\frac{{\rm d}m}{q}.
\end{equation}
If the primaries follow the same power law as the total IMF, $\varphi_{\rm p}(m) \propto m^{-\alpha}$, Eq.~(\ref{eq:phi_s_def}) gives $\varphi_{\rm s}(m) \propto q^{\alpha-1}\varphi_{\rm p}(m)$. Substituting into Eq.~(\ref{eq:phi_sum}), we obtain
\begin{equation}
    \varphi_{\rm p}(m) = \frac{\varphi(m)}{1 + q^{\alpha-1}}, \qquad
    \varphi_{\rm s}(m) = \frac{q^{\alpha-1}}{1 + q^{\alpha-1}}\,\varphi(m).
\end{equation}
For $q = 0.8$ and $\alpha = 2.35$, $q^{\alpha-1} \simeq 0.74$, so that $\simeq 57.5\%$ of the stars of a given mass are primaries and $\simeq42.5\%$ are secondaries. Notice that the shape of the IMF is unchanged as the two components are simply rescaled versions of $\varphi(m)$.

The only exception to this constant splitting is at the upper end of the grid. The most massive primary of the \acr{F23} set has $M_{1, \rm max}=45\ \rm M_\odot$, whose companion has $q\,M_{1, \rm max}= 36\ \rm M_\odot$. Stars with $36 < m/{\rm M_\odot} \leq 45$ cannot be the secondary of any primary within the IMF range, hence we set $\varphi_{\rm p}(m) = \varphi(m)$ and $\varphi_{\rm s}(m)=0$ in this interval, so that the total number of massive stars is the same as when all massive stars are assumed to be single. At the lower end, the secondaries of primaries with $11\leq M_1/\rm M_\odot < 13.75$ have masses $8.8 \leq m/\rm M_\odot < 11$, below the lowest mass of the \acr{F23} grid. Since no yields are available in this range, these stars lock up gas but do not return any ejecta to the ISM. The single-star yields are therefore assigned only to secondaries with $11\leq m/\rm M_\odot\leq 36$. To summarize, we adopt:
\begin{equation}
    \varphi_{\rm p}(m) =
    \begin{cases}
        \varphi(m)/(1 + q^{\alpha-1}) & m \leq q\,M_{1,\rm max}, \\
        \varphi(m)                    & q\,M_{1,\rm max} < m \leq M_{1,\rm max},
    \end{cases}
\end{equation}
\begin{equation}
    \varphi_{\rm s}(m) =
    \begin{cases}
        q^{\alpha-1}\,\varphi(m)/(1+q^{\alpha-1}) & m \leq q\,M_{1,\rm max}, \\
        0                                         & q\,M_{1,\rm max} < m \leq M_{1,\rm max}.
    \end{cases}
\end{equation}

In the GCE model, the two components are then integrated separately, each with its own mass function, yields and lifetimes (see next Sect.).

\subsection{Nucleosynthesis prescriptions}
\label{sec: nucleosynthesis prescriptions}

The models of \acr{LC18} cover the initial masses $\rm M_{\rm ZAMS} = 13, 15, 20, 25, 30, 40, 60, 80$ and $\rm 120\ M_\odot$, four metallicities $\rm [Fe/H] = 0, -1, -2$ and $-3$ ($\rm Z = 1.345\times10^{-2}, 3.236\times10^{-3}, 3.236\times10^{-4}$ and
$\rm 3.236\times10^{-5}$) and three initial equatorial rotational velocities, $\rm v_{rot} = 0, 150$ and $\rm 300\ km\ s^{-1}$, for a total of $108$ models. We adopt their set R, in which the stars with $\rm M_{\rm ZAMS} \leq 25\ M_\odot$ explode with a mass cut calibrated to eject $\rm 0.07\ M_\odot$ of $\rm ^{56}Ni$, after having imposed a prescription for mixing and fallback, whereas the more massive ones collapse entirely into a black hole and contribute to the enrichment of the ISM only through their winds. Since the yields depend on $\rm v_{rot}$, we consider a combined set, obtained by weighting the three tables with the metallicity-dependent distribution of the initial rotational velocities from \citet{Prantzos2018}, that favours fast rotators in the early Galaxy. As illustrated in Tab.~\ref{tab:prog_details}, twelve of the $108$ \acr{LC18} models develop a partial or complete C--O shell merger, and all of them are found among the three lowest masses of the grid. Their K yields exceed those of the non-merger models, with a C--O merger model ejecting roughly 10 times more K than a non-merger star of similar mass (in the range $\rm 13-25\ M_\odot$).

The NuGrid models of \acr{R18} are non-rotating, with initial masses $\rm M_{\rm ZAMS} = 12, 15, 20$ and $\rm 25\ M_\odot$ at five metallicities, $\rm Z = 10^{-4}, 10^{-3}, 6\times10^{-3}, 10^{-2}$ and $2\times10^{-2}$, for a total of $20$ models. This is the only one of the three sets that is complemented by an internally consistent set of low- and intermediate-mass star yields, which we do not use here in order to keep the low- and intermediate-mass contribution identical in all the sets. Four of the models (the $12$, $15$, and $\rm 20\ M_\odot$ models at $\rm Z = 10^{-2}$ and the $\rm 15\ M_\odot$ model at $\rm Z = 2\times10^{-2}$) undergo a C--O shell merger (see Tab.~\ref{tab:prog_details}). They are confined to the two highest metallicities of the grid, which is the reason why this set enhances the production of K and of the other odd-Z elements only towards the high-metallicity end of the Galactic evolution (see \citealp{Ritter2018_ConvReacNuclScClPproIsot}). The effect on the yields is large: the $\rm 15\ M_\odot$ model at $\rm Z = 2\times10^{-2}$, for example, ejects $\rm 2.2\times10^{-3}\ M_\odot$ of K, a factor of $\sim47$ and $\sim22$ more than the 12 and 20$\,\rm M_\odot$ models at the same metallicity, respectively.

\acr{F23} computed a dense grid of MESA models with $\rm M_{\rm ZAMS} = 11 - 45\ M_\odot$ in steps of $\rm 1\ M_\odot$ at solar metallicity, both for single stars ($35$ models) and for stars that are stripped by a companion through Roche-lobe overflow ($31$ models). Their ejecta are given separately for the wind, the Roche-lobe overflow, and the core-collapse phase, and we sum all three contributions. We use this set in three different configurations: i) all massive stars are single, and the single-star yields are applied over the whole IMF ("single stars only"); ii) all massive stars are in binaries and only the primaries contribute, with the stripped-star yields weighted by $\varphi_{\rm p}(m)$ ("donors only"); iii) as ii) but the secondaries also contribute, weighted by $\varphi_{\rm s}(m)$ ("donors plus accretors"; see Sect. \ref{sec: the imf of binary systems}). Since no consistent set of yields is available for the accretors, we assign to the secondaries the single-star yields and lifetimes of the same initial mass, i.e. we treat them as if they evolved in isolation, neglecting the mass they accrete from the donor and the associated rejuvenation (see Sect. \ref{sec: caveats}). 

Since the \acr{F23} grid for donor stars is computed at a single metallicity, the metallicity dependence of the yields of the stripped stars is recovered following the approach of \citet{Pepe2026} and described in \citet{Ma2025}, i.e.:
\begin{equation}
\label{eq: Zdep}
    \begin{aligned}
        M_{\rm ej}(m, Z) &= M_{\rm ej}^{\rm bin}(m) + \left[\,M_{\rm ej}^{\rm bin}(m) - M_{\rm ej}^{\rm single}(m)\,\right] f(Z), \\
        f(Z) &= \max\!\left[-1,\, \min\!\left(0,\, 1.25 \log_{10}(Z/Z_\odot)\right)\right].
    \end{aligned}
\end{equation}
Here, $M_{\rm ej}(m, Z)$ is the mass of a given element ejected by a stripped star of initial mass $m$ and metallicity $Z$, and $M_{\rm ej}^{\rm bin}(m)$ and $M_{\rm ej}^{\rm single}(m)$ are the corresponding ejected masses of the stripped and single star models of the same initial mass at solar metallicity. Finally, $f(Z)$ is a metallicity-dependent weighting factor, chosen so that $M_{\rm ej}(m, Z) = M_{\rm ej}^{\rm single}(m)$ at metallicity $\rm Z \lesssim 0.16\ Z_\odot$, and gradually approach $M_{\rm ej}(m, Z) = M_{\rm ej}^{\rm bin}(m)$ at solar metallicity. 

In this set, the C--O shell mergers occur in the $11$ and $\rm 13\ M_\odot$ stripped models and in the $11$, $13$ and $\rm 18\ M_\odot$ single models. The $11$ and the $\rm 13\ M_\odot$ stripped stars eject $\simeq2-3\times 10^{-3}\ \rm M_\odot$ of K, i.e. $\simeq25$ times more than the models of adjacent mass. 

Besides the massive stars, the model includes the ejecta of low- and intermediate-mass stars (stars with initial masses $1.0 \leq M/\rm M_\odot \leq 8.0$, with yields from the FRUITY database, \citealp{Cristallo2009, Cristallo2011, Cristallo2015}), of Type Ia supernovae (whose rate is computed with the delay-time distribution of \citealp{Greggio2005} and whose yields are taken from model W7 of \citealp{Iwamoto1999}), of novae (following the prescriptions of \citealp{CescuttiMolaro2019}), of neutron-star mergers and of magneto-rotational supernovae (both with prescriptions from \citealp{Molero2023}). All of these prescriptions are kept fixed, and only the massive star yields differ between models.

\section{Results}
\label{sec: results}

\begin{figure}
    \centering
    \includegraphics[width=1\columnwidth]{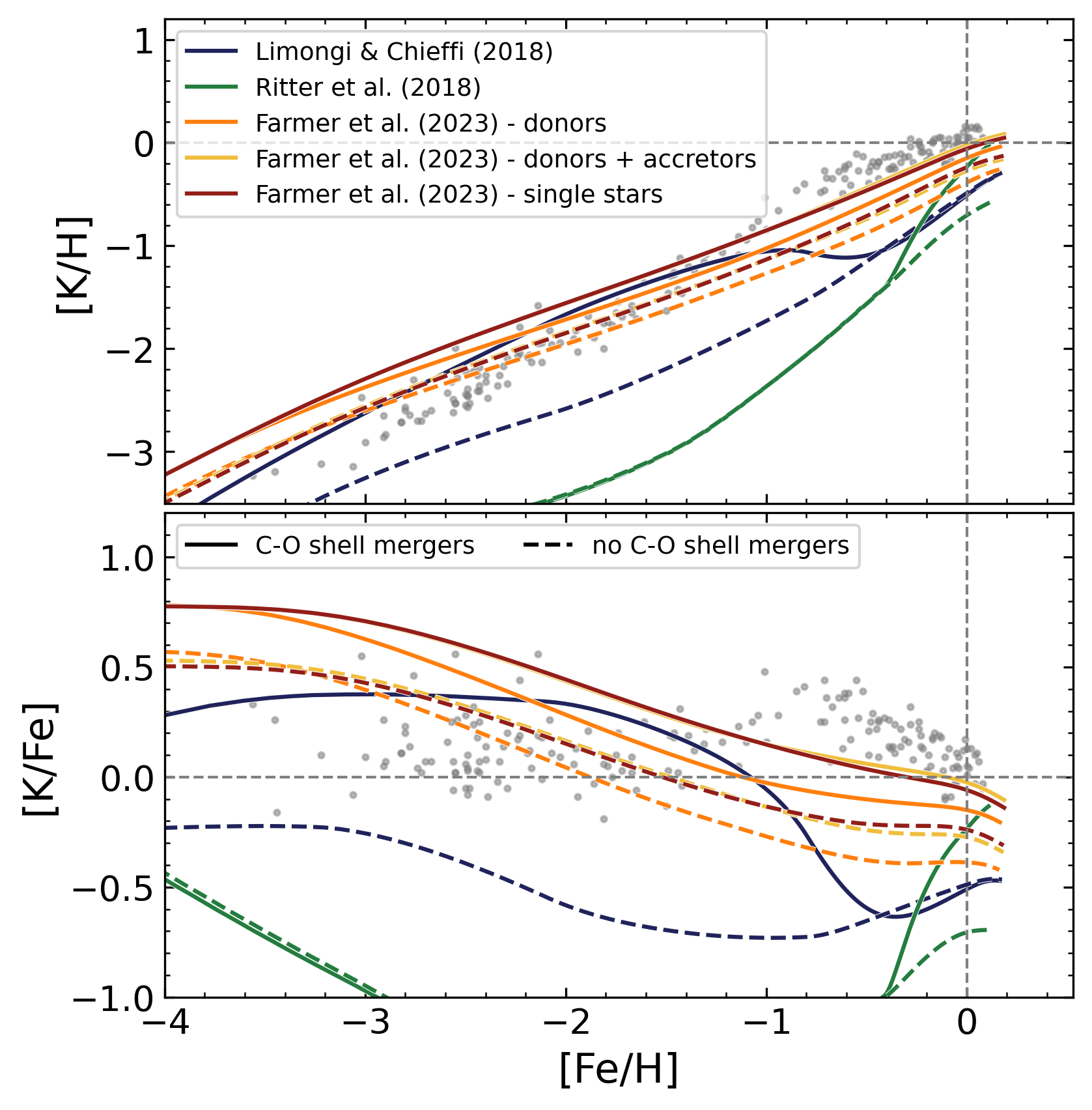}
    \caption{Predicted evolution of K. \emph{Upper panels}: [K/H]; \emph{lower panels}: [K/Fe], both as a function of [Fe/H]. Different colors identify the massive star yield sets adopted: \acr{LC18}, combined over the three rotational velocities with the metallicity-dependent distribution of \citet{Prantzos2018} (blue); \acr{R18} (in green); \acr{F23} with donors only (orange), with donors plus accretors (yellow), and with single stars only (red). Solid lines are computed with the complete yield tables. Dashed lines are the same models in which the yields of the progenitor that undergo a C--O shell merger are replaced by an interpolation over the models without a C--O merger of the same metallicity. Grey points are observational values from \citet{Chen2000} and from \citet{MittalRoederer2025}.}
    \label{fig: K and Sc gce}
\end{figure}

In Fig.~\ref{fig: K and Sc gce}, we show the predicted evolution of K from our GCE model. We provide both the absolute abundance [K/H] and the abundance ratios [K/Fe], both as a function of [Fe/H]. For each set of massive star yields, we computed one model with the complete yield table (i.e., solid lines in Fig.~\ref{fig: K and Sc gce}) and one model where the yields of the progenitors that undergo a C--O shell merger were removed and replaced by an interpolation over the neighbouring progenitors of the same metallicity. All of the other ingredients are identical, so that the only difference between the two models is the contribution (or lack thereof) from C--O merger progenitors.


\begin{figure}
    \centering
    \includegraphics[width=0.9\columnwidth]{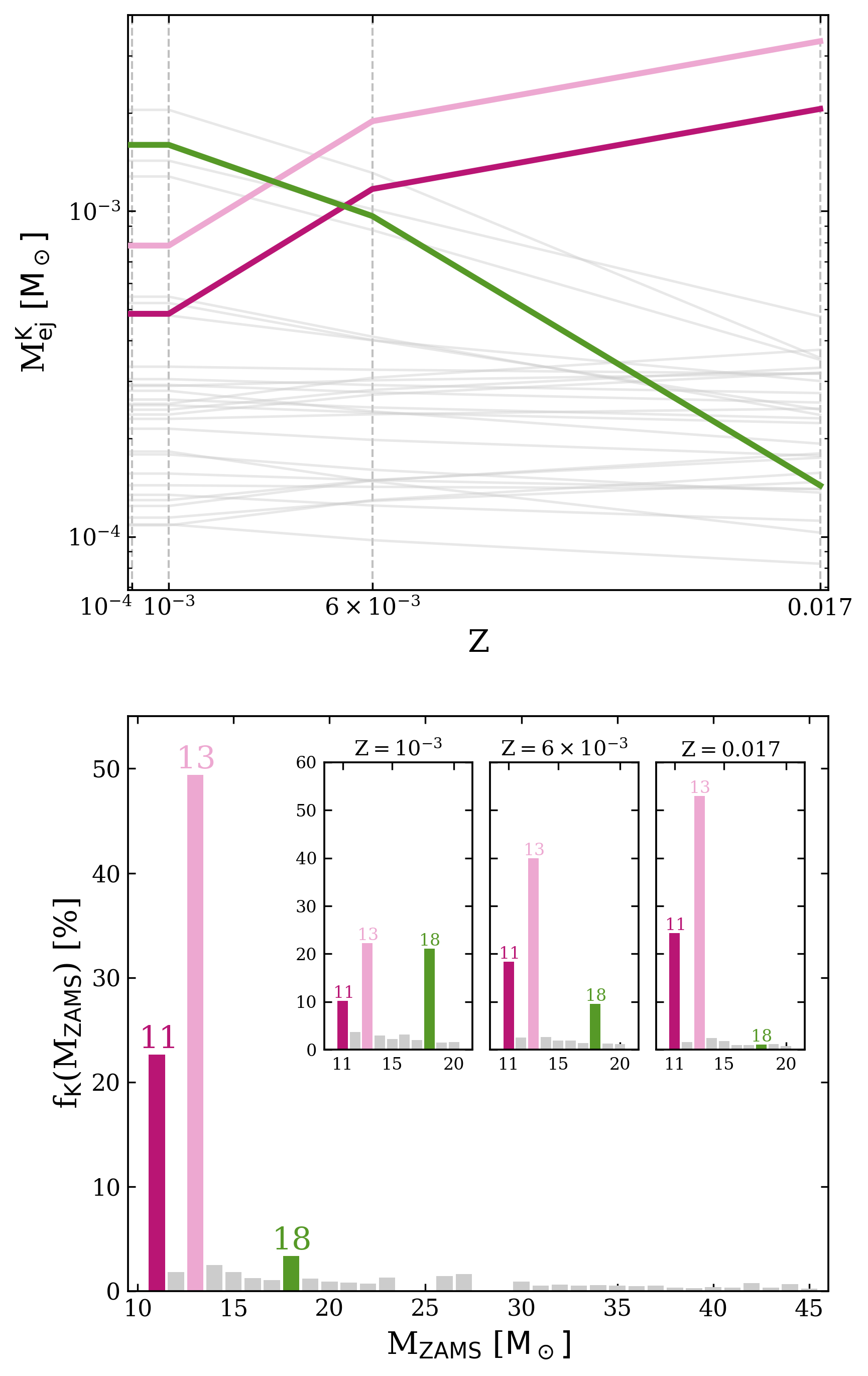}
    \caption{Contribution of the individual progenitor masses to the K production for the donor-only yields of \acr{F23}, with metallicity dependence from Eq.~\ref{eq: Zdep}. \emph{Upper panel}: mass of K ejected by each model as a function of the metallicity $Z$. The $11$, $13$, and $\rm 18\ M_\odot$ models, which undergo a C--O shell merger, are highlighted in colour, all the other masses are shown in grey. \emph{Lower panel}: fraction $f_{\rm K}(M_{\rm p})$ of all the K produced over the history of the Galaxy that comes from each progenitor mass. The insets show the fraction of the K ejected by a single stellar generation contributed by each progenitor mass, at the three metallicities $10^{-3}$, $6\times10^{-3}$ and $0.017$.}
    \label{fig:K_farmer}
\end{figure}

To interpret Fig.~\ref{fig: K and Sc gce}, it is instructive to also look at Fig. \ref{fig:K_farmer} for \acr{F23} (similar figures for the other progenitors can be found in the App. \ref{sec: IMF weighted contribution}, Figs. \ref{fig:K_ritter}, \ref{fig:K_lc} and \ref{fig:KLC}), 
where we decompose the production of K by progenitor mass. In the top panel, we show the total ejected mass of K ($M^{\rm K}_{\rm ej}$) as a function of metallicity, where each line represents a progenitor of a given mass. In the bottom panel we show the fraction of all the K produced over the entire Galactic history coming
from each progenitor mass, obtained by integrating the star formation rate, the IMF and the metallicity-dependent yields over time. For a single stellar generation born at a given $Z$, we also show the computed fraction $f_{\rm K}(M_{\rm p})$ of the K ejected by each progenitor mass, after weighting the yields with the IMF.

\subsection{Binary-stripped models}

From the GCE models that adopt the yields of \acr{F23}, it is clear that the C--O merger progenitors contribute at all metallicities. However, this is a direct consequence of how the set is constructed: the stellar grid is computed at a $Z = Z_\odot$, and through Eq.~(\ref{eq: Zdep}) the yields of a given stellar system at a given metallicity are a combination of its single-star and binary-stripped star yields. Therefore, all of the C--O merger progenitors found at solar metallicity are also present at any other metallicity. As is clearly shown in Fig.~\ref{fig:K_farmer}, in the GCE model with donor stars only, the K budget is dominated by the C--O merger progenitors. In particular, the 11 and $\rm 13\ M_\odot$ models alone supply around $23$ and $50\%$ of all the K produced over the history of the Galaxy, whereas no other mass contributes more than a few per cent. Their relative weight changes with metallicity. At low metallicities, where Eq.~(\ref{eq: Zdep}) returns the single star yields, the K produced by a stellar generation is dominated by the yields of the 11, 13 and $\rm 18\ M_\odot$ models. At solar metallicity, instead, the yields approach those of the stripped stars, with negligible contribution to the K production from the $\rm 18\ M_\odot$ binary-stripped model, which does not have a C--O merger (contrary to its single-star counterpart). Since C--O merger progenitors are present at all metallicities by construction, their effect on the GCE of K is an almost uniform upward shift in [K/Fe] in all three configurations (Fig.~\ref{fig: K and Sc gce}). Accordingly, [K/H] grows almost linearly with [Fe/H], without showing any sudden change of slope. Moreover, [K/Fe] reaches slightly sub-solar values at [Fe/H]$\,=0$, while without the contribution from C--O merger progenitors, it becomes sub-solar already at [Fe/H]$\,\simeq-2$ ("donors" model) and $\simeq-1.5$ (``donors plus accretors" and "single-stars-only" configurations).

For [Fe/H]$\gtrsim-3$, the model that includes donors plus accretors lies above both the donor-only model and the single-stars-only model, with and without C--O merger progenitors. Below this metallicity, instead, the three models coincide. At low [Fe/H] both K and Fe come only from core-collapse supernovae. As [Fe/H] increases, however, Type Ia SNe start to contribute to the Fe production. As a consequence, since in the donors model only the primaries explode, the Fe from Type Ia SNe lowers the [K/Fe] more than in the model that also includes the accretors (in which both the primaries and secondaries explode) and in the single stars model (in which all of the massive stars explode). The close agreement between the single-stars-only and the donors plus accretors models shows that binary interaction by itself is not the main driver of the K production, which is instead set by the presence of C--O merger progenitors.


\subsection{Single-star models}

The two single-star sets considered here, \acr{LC18} and \acr{R18}, extend to lower metallicities and therefore the frequency of C--O merger progenitors varies significantly with $Z$ since it is not imposed by construction, as it is instead for the set from \acr{F23}. In particular, in the \acr{LC18} models, the C--O merger progenitors are frequent in the early Galaxy and become much more rare close to solar metallicity. In the \acr{R18} models, on the other hand, C--O merger progenitors are absent in the early Galaxy and more frequent closer to solar metallicity. In both cases, the metallicity range over which [K/H] and [K/Fe] are enhanced coincides with the range over which C--O merger progenitors are present in the grid. 

Without C--O merger progenitors, the \acr{LC18} models predict a sub-solar [K/Fe] at all metallicities, clearly below the stellar observations from \citet{Chen2000} and \citet{MittalRoederer2025}. Including the C--O merger progenitors raises [K/Fe] to super-solar values ($\simeq 0.25\ \rm dex$) for $\rm [Fe/H] \lesssim -1.5$, producing a low-metallicity plateau that reproduces the shape of the observed trend, although lying above the observed values. Such enhancement fades quickly between [Fe/H]$\simeq-1.5$ and $0.5\ \rm dex$. The two models, with and without C--O merger progenitors, then converge to sub-solar values of [Fe/H], lying well below the data (see Fig.~\ref{fig:KLC} for more details). At the two lowest metallicities of the grid, the rotating (and, at the lowest Z, also the non-rotating) $\rm 15\ M_\odot$ models undergo a C--O shell merger, and they alone supply most of the K of their generation. At $\rm Z = 3\times10^{-3}$ the C--O merger model is the $\rm 13\ M_\odot$ star rotating at $\rm 150\ km\ s^{-1}$, which again dominates its stellar generation. At these metallicities, the adopted rotational velocity distribution assigns a large weight to the rotating models, which are precisely those with a C--O merger. At solar metallicity, instead, the only C--O merger models are the $\rm 20\ M_\odot$ star with $\rm v_{rot} = 150\ km\ s^{-1}$ and the $\rm 13\ M_\odot$ star with $\rm v_{rot} = 300\ km\ s^{-1}$. However, the distribution is dominated by non-rotating stars, among which no C--O merger occurs at this metallicity. In the solar metallicity panel of the non-rotating stars of Fig.~\ref{fig:KLC}, K is shared among the $\rm 13-25\ M_\odot$ progenitors without a single dominant mass. The drop of the solid curve at [Fe/H]$\simeq-1\ \rm dex$ therefore marks the transition from a K budget controlled by a few C--O merger rotators to one controlled by the stars without C--O shell mergers, further enhanced by the onset of Fe production by Type Ia supernovae. 

The \acr{R18} models show the opposite trend. Below [Fe/H]$\simeq -0.5\ \rm dex$, they predict a very low [K/H] and [K/Fe], and the solid and dashed curves coincide, since there are no C--O merger progenitors in the grid below $\rm Z = 10^{-2}$. Above this metallicity, both [K/H] and [K/Fe] rise steeply and approach the solar value only when the C--O merger progenitors are included, while they remain well below it without them. This is even clearer if one looks at the K yields from the $12$, $15$ and $\rm 20\ M_\odot$ models (see Fig.~\ref{fig:K_ritter}). There is a sharp increase of more than an order of magnitude from $\rm Z = 6 \times 10^{-3}$ to  $\rm Z = 10^{-2}$ because those three models only experience a C--O merger at $\rm Z = 10^{-2}$ and not at $\rm Z = 6 \times 10^{-3}$. The $\rm 15\ M_\odot$ model, which experiences a C--O merger at both the highest metallicities, supplies the large majority of the K produced over the history of the Galaxy. It should be stressed that the fact that the $\rm 15\ M_\odot$ model also dominates the relative budget at the lowest metallicities of the grid is not in contradiction with this picture, as at those metallicities the absolute K yields are orders of magnitude smaller, and therefore those stars contribute negligibly to the overall Galactic K production. For a detailed analysis of this set, we refer the reader to the work of \citet{Ritter2018_ConvReacNuclScClPproIsot} and \acr{R18}, who first identified the C--O shell merger progenitors within their grid as the origin of the enhanced production of K and of the other odd-Z elements at high metallicity.

Overall, the two model sets based on single-star yields portray a relatively simple qualitative picture. Neither of them reproduces the K observed in Galactic stars without the contribution from C--O merger progenitors, and in both of them the metallicity range over which the GCE model approaches the observations is set by the position of C--O merger progenitors in the $(M_{\rm p}, Z)$ plane, rather than by the details of the chemical evolution model. The Galactic evolution of K is therefore an indication of how frequent C--O merger progenitors are in terms of mass and, even more importantly, metallicity. As discussed in Sect. \ref{sec: C--O shell mergers}, the nucleosynthesis of odd-Z elements in a C--O shell merger is primary \citep{Roberti2025_SPArBurnProtCaptPoweCarbShel}, i.e., it only depends on the C, Ne, and O synthesized during the life of the star, and not on the pre-existing abundance of odd-Z elements itself. From this it follows that it is the frequency of C--O merger progenitors as a function of metallicity that shapes the Galactic evolution of K, rather than the amount of K produced in each event. 

If this is correct, it would imply that to reproduce the linear increase of [K/H] as a function of [Fe/H] observed in stars across metallicities, C--O shell mergers should occur at all metallicities. A C--O merger contribution confined to a limited metallicity interval, as in the \acr{LC18} and \acr{R18} grids considered here, results in a sudden change of slope in the [K/H] vs. [Fe/H] trend, contrary to what is observed in stellar spectroscopy. In addition, it is important to underline that since C--O merger progenitors are usually limited to lower-mass stars ($\rm M_{\rm ZAMS} \simeq 11-20\ M_\odot$), when weighted with an IMF, even one or two of them can have a significant impact.

As a consequence, the observed chemical evolution of K is a potentially powerful tool to constrain how often C--O shell mergers occur in a stellar population as a function of metallicity. 

\subsection{Discussion}
\label{sec: caveats}

To put these results in context, this section critically discusses several limitations of the models and points to where they can be improved. In the case of the single-star sets, our results rely on coarse initial mass grids for the progenitor models. In the \acr{R18} and \acr{LC18} sets, a C--O shell merger is a single $(M_{\rm p}, Z)$ point. Since the GCE model interpolates the yields in mass and metallicity, the C--O merger contribution may spread over a wider range of mass and metallicity than it otherwise would with a finer progenitor grid. For the same reason, a coarse grid might miss selected masses where C--O shell mergers occur, hence overestimating the yields from the stars without a C--O merger. On top of that, in the \acr{LC18} set, all of the C--O merger models, except for one, are rotating stars, and therefore the contribution of the C--O merger progenitors to the Galactic K depends directly on the adopted distribution of initial rotational velocities, which is itself calibrated on the abundances of other elements (e.g., \citealp{Prantzos2018, Romano2019, Rizzuti2021, Molero2024}). This implies that not only a shift in the C--O merger metallicity window, but also a different weight of the rotating models, would move the metallicity range over which [K/H] and [K/Fe] are enhanced. The main result, however, is independent of these details:  the K excess predicted by the models is carried by the C--O merger, and the Galactic evolution of K follows their location in the $(M_{\rm p}, Z)$ plane. 

Concerning the models with the \acr{F23} sets, it must be noted that thanks to its finer initial mass grid, each C--O merger occupies a single $\rm 1\ M_\odot$ bin rather than being interpolated over several solar masses, and yet the C--O merger progenitors are still the main driver of the Galactic K enrichment. On the other hand, whether these progenitors also experience C--O shell mergers at low metallicity is an open question, since in this study the C--O merger frequency as a function of $Z$ follows from Eq.~(\ref{eq: Zdep}) and not from self-consistent stellar evolution models of low-metallicity stars. The near-linear [K/H] obtained with this set should then be regarded as a test of what a C--O merger contribution present at all metallicities produces, rather than as a prediction. Another caveat with the use of yields from binary stars in GCE models is the treatment of the accretors. While the ejecta of the donors are the sum of the wind, the Roche-lobe overflow, and the core-collapse contributions, when the accretors are included, they are assigned the yields of single stars of their initial mass, multiplied by the mass ratio $q=M_2/M_1=0.8$. This neglects the mass that they gain from the donor, as well as the associated rejuvenation. Overall, as also pointed out by \citet{Pepe2026}, the GCE predictions obtained with binary star yields should be interpreted with caution, and mostly in terms of the relative differences between models, by comparing results that differ in a single ingredient, as done here with and without the C--O shell mergers.

Finally, the Galactic evolution of K depends not only on which progenitors undergo a C--O shell merger, but also on which of them explode. In our models, explodability is fixed in each set to what it has been assumed in the original work. For \acr{F23} and \acr{R18}, all massive stars in their grid explode, whereas for \acr{LC18}, only stars below $25\ \rm M_\odot$ explode. The effect of the explodability on the curves shown in Fig. \ref{fig: K and Sc gce} is already visible at [Fe/H]$\,\lesssim-3$. There, the \acr{F23} GCE model without C--O merger progenitors predicts a higher K production than the \acr{LC18} model with C--O merger progenitors, which seems to contradict what was shown so far in this work. However, 
above $\rm 25\ M_\odot$, all the \acr{F23} progenitors explode and eject large amounts of K ($\gtrsim2\times10^{-4}\ \rm M_\odot$ per progenitor, with some of the most massive ejecting around $1.5\times10^{-3}\ \rm M_\odot$). On the other hand, in the set R of \acr{LC18} adopted here, those same stars collapse into a black hole and contribute to the enrichment of the ISM only through their winds. In the \acr{F23} models, these stars, being the first to explode, raise the overall K production and enrich the ISM from very early times, whereas in the \acr{LC18} models, the first supernovae come from K-poor progenitors of $\simeq 20-25\ \rm M_\odot$ without a C--O merger, and the K production rises only later, once the lower-mass stars with a C--O merger (around $15\ \rm M_\odot$, see Tab.~\ref{tab:prog_details}) start exploding.

Self-consistent neutrino-driven explosions, however, predict a much less regular pattern of successful and failed explosions. \citet{Boccioli2026_ExplMattHowRealNdriExplChan} found that several of the lower-mass progenitors ($\sim12-15\ \rm M_\odot$) fail to explode, while some much more massive ones explode. \citet{Pollin2026} showed that, for the same \acr{F23} progenitors adopted here, the outcome depends strongly on how the explosion is modelled. Only at $\rm 11\ M_\odot$ do both the single and the binary-stripped progenitors explode in all the treatments they compared. The $\rm 13\ M_\odot$ single C--O merger progenitor, for example, explodes in their simulations, but fails to explode both in \citet{Boccioli2026_ExplMattHowRealNdriExplChan} and according to the explodability criterion of \citet{Ertl2016_TwopCritClasExplMassStarNdri}. Since a few low-mass C--O merger progenitors supply most of the Galactic K in our models, the failure of even one of them would substantially lower the predicted [K/Fe] and [K/H]. The effect of explodability on the chemical evolution of both light and heavy elements will be the subject of a future, more extended study.
 
\section{Conclusions}
\label{sec: conclusions}

In this work, we have investigated the contribution of the C--O shell mergers in massive stars to the Galactic chemical evolution (GCE) of K, by selecting the C--O merger progenitors in three sets of massive star yields and by comparing GCE models computed with and without their contribution. Our main conclusions are the following:

\begin{itemize}
    \item Without progenitors that experienced C--O shell merger, none of the GCE models can reproduce the observed evolution of K, and they predict a sub-solar [K/Fe] over most of the metallicity range.

    \item When the C--O merger progenitors are included, [K/Fe] is enhanced only over the metallicity range in which C--O merger progenitors are present in the grid: at [Fe/H]$\lesssim -1\ \rm dex$ for the rotating models of \acr{LC18}, and close to solar metallicity for the models of \acr{R18}. The Galactic evolution of K thus traces the location of C--O merger progenitors in the mass-metallicity plane.

    \item In the \acr{F23} models, a few C--O merger progenitors (mainly the 11 and $\rm 13\ M_\odot$, both stripped and single stars) supply most of the Galactic K. This is also a consequence of Eq.~(\ref{eq: Zdep}), which extends the C--O merger progenitors found at solar metallicity to all $Z$. The effect is an almost uniform upward shift of [K/Fe], resulting in a linear [K/H] vs. [Fe/H] relation. At [Fe/H]$\,\lesssim-3$, the production of K is high even without C--O merger progenitors, because all massive stars up to $\rm 45\ M_\odot$ explode and eject large amounts of K. We remind that, at these metallicities, the progenitors are the single-star ones.

    \item Since the odd-Z production in a C--O shell merger is essentially primary, the observed [K/H], which increases almost linearly with [Fe/H], can be reproduced only if C--O shell mergers occur with a certain frequency at all metallicities. A C--O merger contribution confined to a limited metallicity range instead causes a change of slope in the [K/H] vs. [Fe/H] relation.

    \item The use of binary star yields in GCE models relies on several assumptions concerning their metallicity dependence, the binary fraction, the mass-ratio distribution and the treatment of the accretors. Because a C--O shell merger either occurs or does not, causing a very sudden change in yields, interpolating (or extrapolating) in mass and metallicity across C--O merger and non-C--O merger progenitors is particularly uncertain.
\end{itemize}

Beyond the points listed above, robust conclusions regarding the occurrence of C--O shell mergers as a function of mass, metallicity, and rotation, are not possible at this stage. However, with more detailed numerical studies of the late phases of stellar evolution, we will be able to learn more about how they develop and affect the nucleosynthesis and the chemical evolution of elements. 


%

\begin{acknowledgements}
The authors thank Joshua Issa, Francesca Matteucci, Emanuele Pepe, Federico Rizzuti, and Lorenzo Roberti for the useful discussions. LB would like to thank the N3AS center for its hospitality and support.

This work was supported by the Deutsche Forschungsgemeinschaft (DFG, German Research Foundation) – Project-ID 279384907 – SFB 1245.
This work was supported by the U.S. Department of Energy under Grant No. DE-SC0004658 and SciDAC grant, DE-SC0024388.

MM acknowledges the Momentum MSCA Program, co-funded by the European Commission through the HORIZON-MSCA-2023-COFUND program and the Secretariat of the Hungarian Academy of Sciences (MTA), Grant Agreement No. 101179854.
\end{acknowledgements}
\bibliographystyle{bibtex/aa} 
\bibliography{Bibliography,MyLibrary.bib} 

\begin{appendix}
\nolinenumbers

\section{IMF-weighted contribution to K}
\label{sec: IMF weighted contribution}

\begin{figure}
    \centering
    \includegraphics[width=0.9\columnwidth]{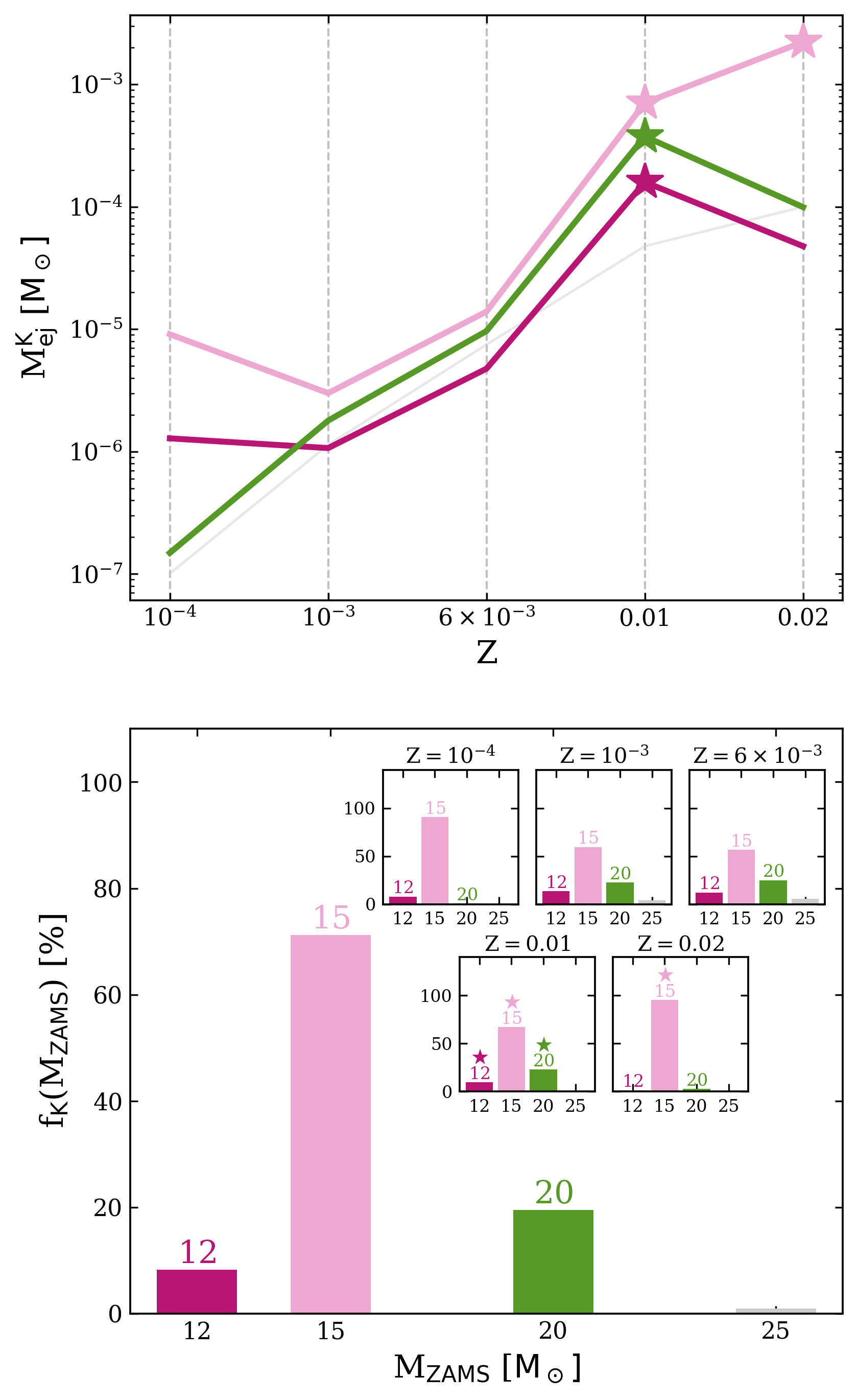}
    \caption{Same as Fig.~\ref{fig:K_farmer} but for the massive star yields set of \acr{R18}.}
    \label{fig:K_ritter}
\end{figure}

\begin{figure}
    \centering
    \includegraphics[width=0.9\columnwidth]{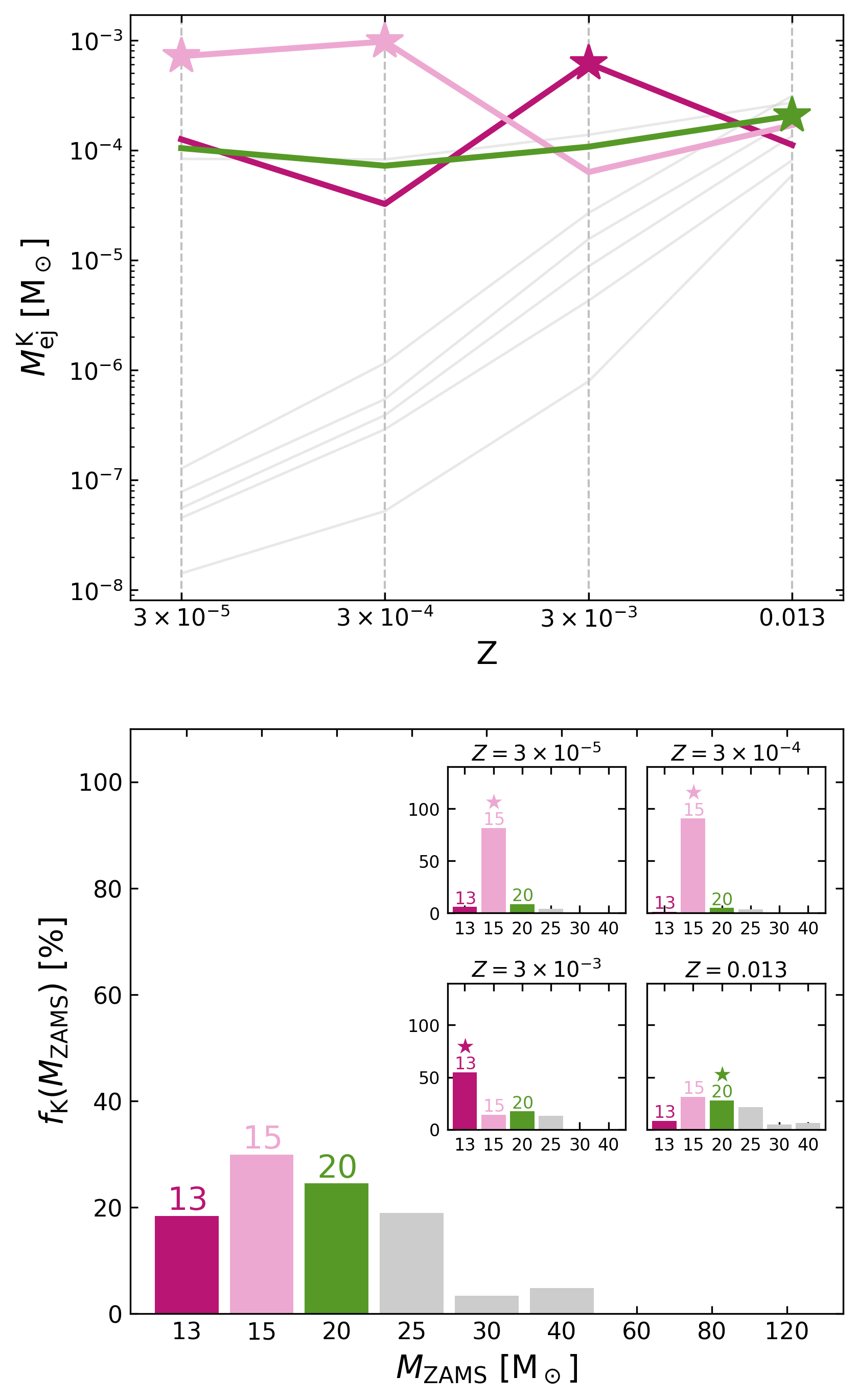}
    \caption{Same as Fig.~\ref{fig:K_farmer} but for the massive star yields set of \acr{LC18}, weighted by the adopted initial rotational velocity distribution. Here a star marks a mass--metallicity combination where at least one of the rotating models that enter the weighted average (with non-zero weight) undergoes a C--O shell merger. See Fig.~\ref{fig:KLC} for the contribution of each rotational velocity separately.}
    \label{fig:K_lc}
\end{figure}

\begin{figure*}
    \centering
    \includegraphics[width=1\textwidth]{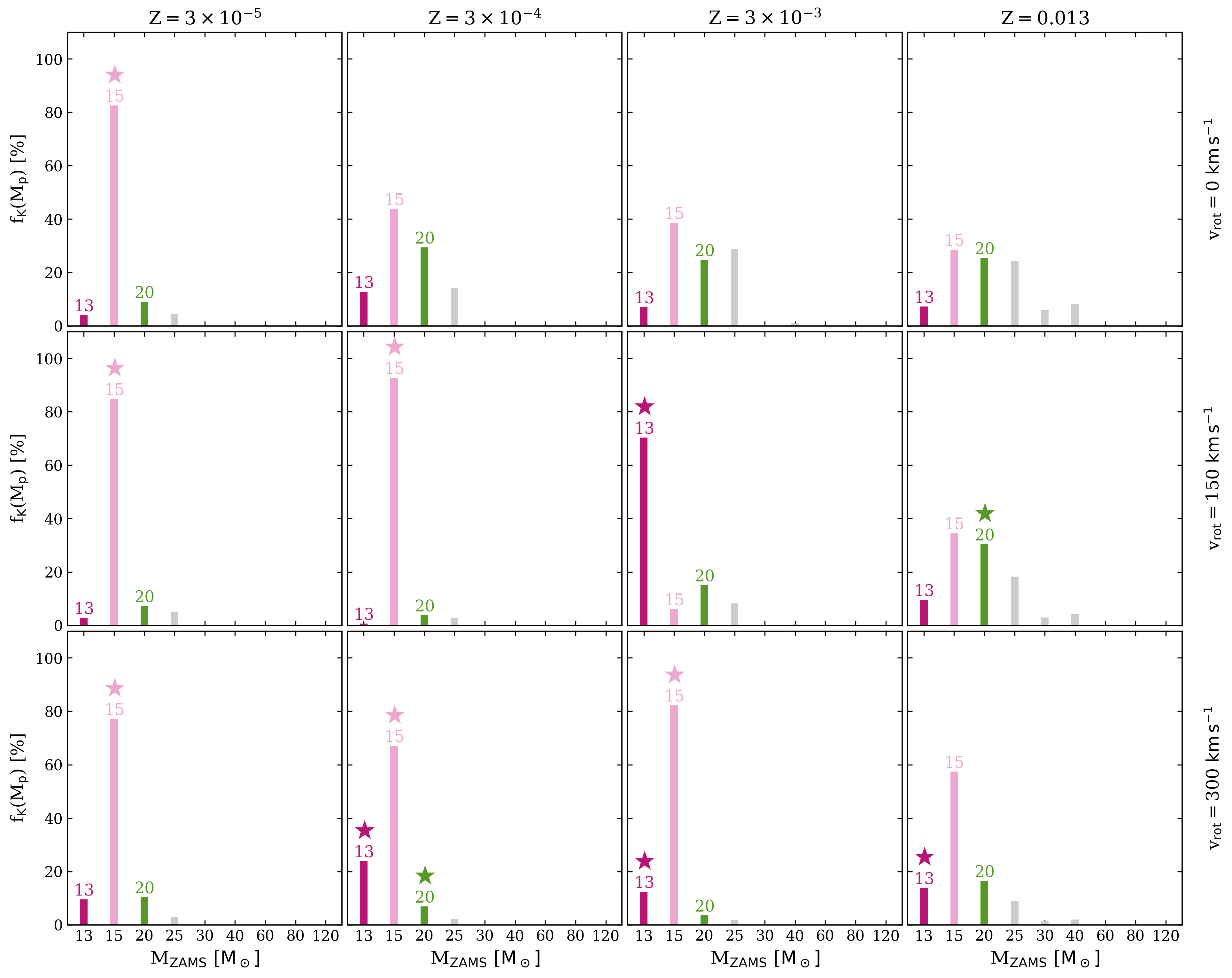}
    \caption{Same as in the lower panels of Fig.~\ref{fig:K_farmer}, but for the set R of \acr{LC18}. Columns correspond to the four metallicities of the grid and rows to the three initial rotational velocities, each shown separately, i.e. without the rotational velocity distribution. Progenitors which undergo a C--O shell merger in at least on $(v_{\rm rot}, Z)$ combination, are highlighted in color, while stars mark the models that actually experience a C--O merger (see Tab.~\ref{tab:prog_details}). Notice that, in some cases, some C--O merger progenitors contribute less Potassium than other stars. This can be seen, for example, in the highest rotation and metallicity, i.e., the bottom right corner of the figure. It should be highlighted that this is a direct effect of how the explosion is tuned in \acr{LC18}. Due to their bomb prescriptions, high-compactness stars (often the ones with high initial rotation) experience significant explosive nucleosynthesis that extends quite far in mass, causing explosive oxygen burning to produce large amounts of K. With more realistic explosions \citep{Boccioli2026_ExplMattHowRealNdriExplChan}, one would not expect this to be the case. }
    \label{fig:KLC}
\end{figure*}

Figs. \ref{fig:K_ritter}, \ref{fig:K_lc} and \ref{fig:KLC} show, for the \acr{R18} and \acr{LC18} sets, the decomposition by progenitor mass discussed in Sect. \ref{sec: results}.

\section{Development of C--O shell mergers}
\label{sec: Kippenhahn diagrams}

\begin{figure*}
    \centering

    \includegraphics[width=0.5\textwidth]{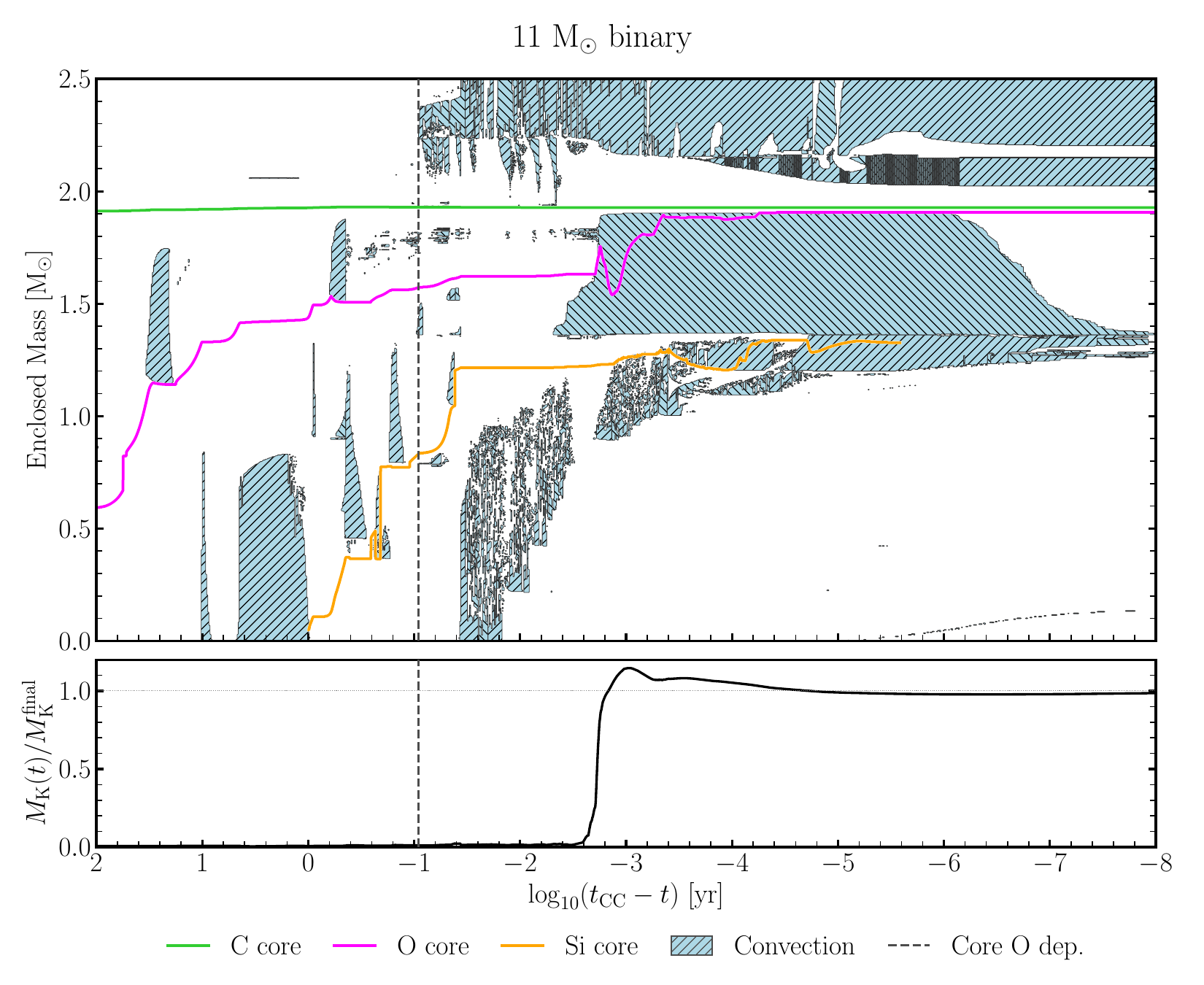}
    \includegraphics[width=0.5\textwidth]{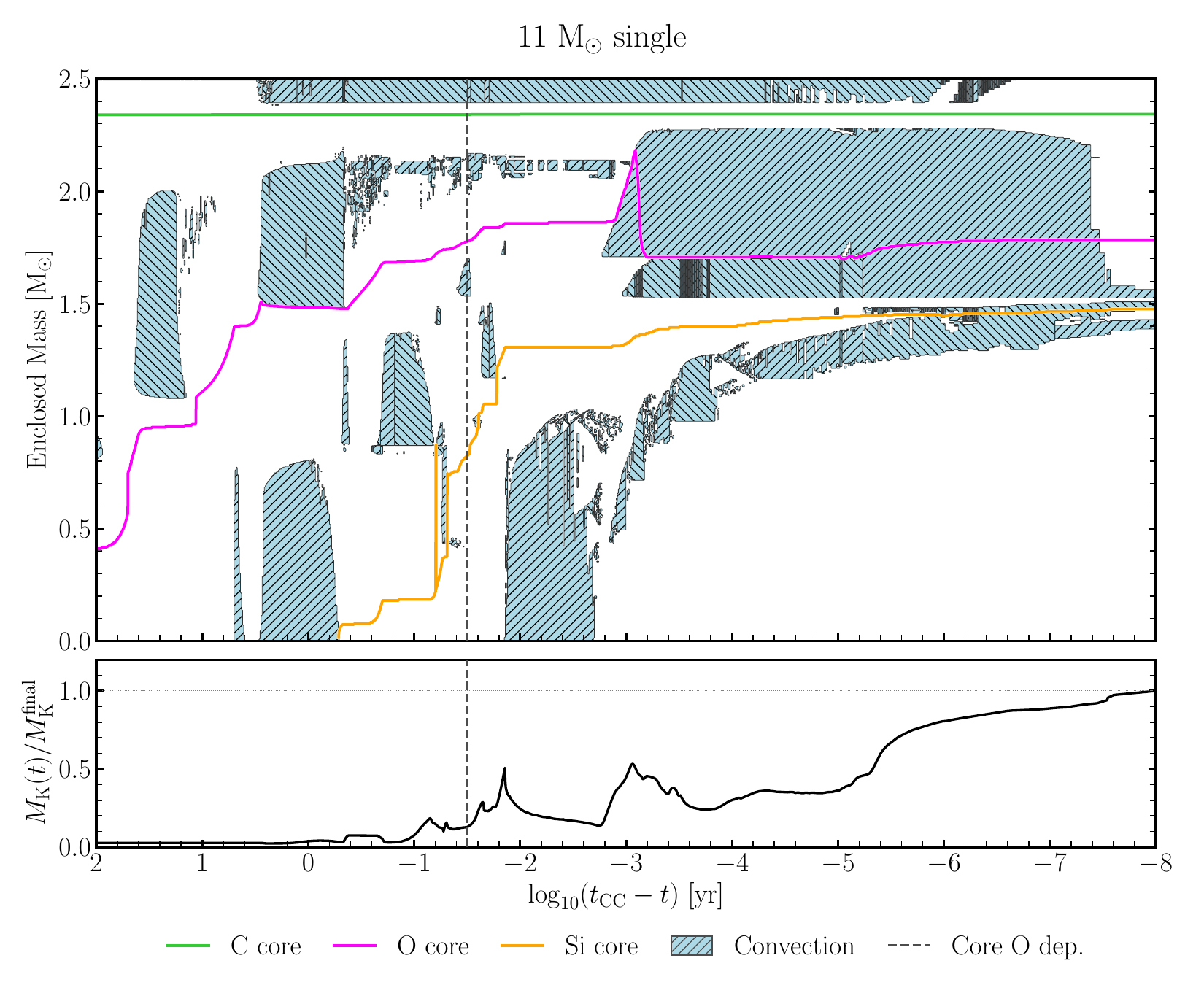}
    
    \caption{Kippenhahn diagram of the $11\ \rm M_\odot$ binary-stripped star (left) and of the $11\ \rm M_\odot$ single star (right) from \acr{F23}, generated using the MESA history files from \acr{F23} found on Zenodo: \url{https://zenodo.org/records/5929871}. The x-axis shows the logarithmic time left before core-collapse, where $t_{\rm CC}$ is the age at the end of the stellar evolution calculation. The convective regions are shown in dashed blue. The vertical dashed line marks the core O depletion, defined as the time when the central $^{16}$O mass fraction drops below $10^{-4}$. The C core, O core, and Si core lines are the core boundaries as computed by \texttt{MESA}, i.e. the outermost mass coordinates where the mass fraction of the preceding fuel drops below 0.01: $^4$He for the C core, $^{12}$C for the O core, and $^{16}$O for the Si core. The bottom panels show the total mass of K ($^{39}$K+$^{40}$K+$^{41}$K) in the entire star, normalized to its value at the end of the calculation.}
    \label{fig: Kipp_11}
\end{figure*}

To illustrate the development of a C--O merger, we show Kippenhahn diagrams of the $11\ \rm M_\odot$ binary-stripped and single stars from \acr{F23} in Fig.~\ref{fig: Kipp_11}. Similar diagrams illustrating the development of a C--O shell merger can also be found in \citet{Ritter2018_NuGrStelDataSetIIStelYiel}, \citet{Rizzuti2024_ShelMergLateStagMassStarEvol}, and \citet{Roberti2025_SPArBurnProtCaptPoweCarbShel}. 

In Kippenhahn diagrams it sometimes happens that convective regions look like they overlap visually, even though they are separated by a very thin mass shell. Therefore, we hash convective regions that are separated by thin mass shells with dashes in opposite directions. This is particularly relevant for the case of C--O shell mergers since in the 11 $\rm M_\odot$ single star, from $\sim 10^{-3.5}$~yrs to $-10^{-5}$~yrs before core-collapse, there are two distinct convective regions. The first one is a an O-burning shell sitting at a mass coordinate of $\sim 1.5\ \rm M_\odot$, with another convective region on top sitting at a mass coordinate of $\sim 1.7\ \rm M_\odot$. The start of the C--O merger is very clear in this star: at $\sim 10^{-3}$~yrs before core-collapse, a convective region develops, quickly moving the O-core line to larger masses. This signals that \isotope[12]{C} from the C-shell is being ingested into the O-region and burned (see the definition of the O-core line in Figure~\ref{fig: Kipp_11}). Then, the O-core line quickly goes back and settles at the bottom of the convective C-shell, signalling that after mixing the \isotope[12]{C} settles back to an equilibrium value above 0.01. K production follows a similar trend, with a temporary peak in production when the \isotope[12]{C} is ingested, and settling back down to a lower value (but larger than before the ingestion event). From this, one can infer that C is not burned anymore, and hence neither is O. 

The most significant K production occurs around $10^{-5}$~yrs before collapse, when the two convective zones merge, allowing material in the C-shell to reach the hot O-burning temperatures and trigger the typical C--O merger nucleosynthetic event. However, considering that the size of the convective zone is $\sim 2 \times 10^9$~cm, and using typical convective speeds around $10^{7}$~cm/s, the typical convective timescale at this stage is $\sim 5 \times 10^{-6}$~yrs, i.e. only about two convective overturns occur before core-collapse. This explains why in the final abundances shown in Fig.~\ref{fig:StarStructures}, the mass fractions of most elements are still not flat, as convection has not yet fully mixed the region. 

On the contrary, for the 11 $\rm M_\odot$ binary star, when \isotope[12]{C} is ingested into the zone below $10^{-3}$~yrs before core-collapse, there is only one large convective region going down to the bottom of the O-burning shell. Therefore, hot O-burning temperatures are reached immediately, and hundreds of convective overturns can occur before core-collapse, which is why \isotope[12]{C} is quickly depleted, and the O- and C-core coincide at the end of the stellar evolution calculation. In the case of the 11 $\rm M_\odot$ single star, only at the very end \isotope[12]{C} starts being efficiently burned, causing the O-core line to rise.

It is worth pointing out that, at the pre-collapse stage, K is also produced in the Si shell (see Fig.~\ref{fig:StarStructures}), and that will therefore also be counted in the total mass of K shown in the bottom panels of Figure \ref{fig: Kipp_11}. However, when a C--O merger occurs, the K in the C--O shell is much more abundant: about 80\% and 97\% of the total in the case of the 11 $\rm M_\odot$ single and binary stars, respectively. Finally, the K produced in the Si shell is destroyed regardless during the explosion, and that entire region is completely reprocessed by explosive burning, hence destroying any pre-supernova imprint, including K.

\end{appendix}
\end{document}